%
%
%
\magnification=\magstep1

\overfullrule=0pt
\hsize=6true in
\hoffset=0.45true in
\baselineskip=23pt
\parskip=8pt
\raggedbottom
\line{\bf\hfill}
\vskip 1cm
\centerline {\bf SMALL-SCALE-FIELD DYNAMO}

\vskip 1cm 
\centerline {A.Gruzinov, S.Cowley$^{\dagger }$, R.Sudan$^{\dagger \dagger }$} 

\bigskip  
\centerline {Institute for Advanced Study, School of Natural Sciences}
\centerline {Princeton, NJ 08540}

\centerline {$^{\dagger }$ Department of Physics, University of California, Los
Angeles }
\centerline {405 Hilgard Avenue, Los Angeles, CA 90024-1547}

\centerline {$^{\dagger \dagger}$ Laboratory of Plasma Studies, Cornell
University}
\centerline {Ithaca, NY 14853-7501}
\vskip 0.2in
\centerline {October 9, 1996}
\vskip 1cm
{\bf Abstract.} Generation of magnetic field energy, without mean field
generation, is studied. Isotropic mirror-symmetric turbulence of a
conducting fluid amplifies the energy of small-scale magnetic perturbations if 
the magnetic Reynolds number is high, and the dimensionality of space $d$ 
satisfies $2.103<d<8.765$. The result does not depend on the model of 
turbulence, incompressibility and isotropy being the 
only requirements.

91.25.C.

\vfill
\eject

Solar and galactic magnetic fields are generated by turbulent motions of 
their electrically conducting quasifluid constituents. Generation of 
large-scale or mean magnetic fields is described by the kinematic dynamo 
theory of Steenbeck, Krause and Radler [1]. Small-scale magnetic fields can 
come from the large-scale fields because of turbulent stretching and twisting 
and play an important role in saturation of the dynamo generated mean fields 
[2-4]. It was noted long ago [5-7] that small-scale fields (SSF) can also be 
generated in the absence of mean fields. For example, mirror-symmetric 
turbulence is incapable of the mean field dynamo, but can amplify SSF 
perturbations. A nice qualitative picture of this process was given by 
Moffatt [8]. When the magnetic Reynolds number is large, the magnetic field is 
frozen into the fluid. The field magnitude grows as the infinitesimal line element,
 that is exponentially. However, the characteristic wavenumber of the magnetic 
perturbation also grows exponentially and finally reaches the dissipation scale.
 It is a delicate question which one of these effects wins on the average.

In many astrophysically interesting cases the viscosity is much larger than the
resistivity, and the turbulent velocities cut off at a scale much larger than
the scale at which resistivity destroys magnetic field. The SSF dynamo exists
on scales between the viscous and resistive cut offs. In high Reynolds number
turbulence the SSF dynamo growth rate exceeds the growth rate of the mean field
dynamo by a factor of $Re^{1/2}$. Thus, if SSF dynamo does operate it should be important for the large-scale dynamo 
saturation [3]. Also of interest are the consequences of the SSF dynamo for 
the applicability of the kinematic mean field dynamo as discussed by Kulsrud 
and Anderson [9]. In this letter, we show that the SSF dynamo indeed operates 
for spatial dimension $d$ in the interval $2.103<d<8.765$. The result is 
model-independent, in particular turbulence is not assumed to be 
$\delta $-correlated in time.

To get this result, we use a zero-dimensional
representation of the induction equation. Consider a small element
of fluid with dimensions much smaller than the smallest scale of the turbulent
flow, but much larger than the scale of the magnetic field. We may then
approximate the velocity field of the fluid element by a Taylor expansion
around an arbitrary center ${\bf r}_0$. Thus, $v_a=\dot{r}_{0a}+V_{ab}
(r_b-r_{0b})+...$. Here $V$ is the rate of strain tensor moving with the
fluid element, $V_{ab}=\partial _av_b$. Note that the rate of strain tensor is
dominated by the smallest scales close to the viscous cut off [9]. We assume 
incompressibility, $trV=0$. Then the magnetic field evolution is given by 
equations for the field $B$ and the 
wavenumber $k$ [10]:
$$\dot{B}=VB, \eqno(1)$$
$$\dot{k}=-V^tk,\eqno(2)$$
here $V^t$ is the transpose of $V$; the molecular magnetic diffusivity will 
be taken into account later. Equations (1),(2) are applicable for SSF dynamics 
only, at large length scales they are invalid. More precisely, (1), (2) are always exact, being just equations for co- and contravectors. However, if the length scale of the magnetic field becomes comparable to the length scale of the turbulent velocity field, it takes infinite number of such equations to describe the time evolution of the magnetic field. This infinite system of equations is equivalent to the standard induction equation $\partial_tB=\nabla \times (v\times B)$. Formal solution to this linear equation can be written, but we do not know how to average the resulting magnetic energy. The linear (kinematic) dynamo is a non-trivial problem.

 The system (1),(2) is much 
simpler than the standard induction equation, all the relevant 
properties of turbulence being concentrated in the stretching tensor $V$. We 
wish to know whether the energy of the SSF grows or decays for a random 
stretching rate $V$ with given statistics.

As an introduction to the problem, consider the Gaussian $\delta$-correlated 
turbulence. Then turbulence is specified by the pair correlator of the 
stretching tensor. From isotropy and incompressibility,
$$<V_{ab}V_{cd}>=2\gamma \delta (t)[(d+1)\delta _{ac}\delta _{bd}-\delta _{ab}
\delta _{cd}-\delta _{ad}\delta _{bc}],\eqno(3)$$
where $\gamma$ is a constant with dimension of frequency. The physical 
meaning of $\gamma$ is the characteristic stretching rate, $\gamma$ is
determined by the smallest eddies.  Let $P(k,B)$ be 
the probability distribution for the pairs $(k,B)$. Since $V$ of (1),(2) is 
$\delta$-correlated in time, one can derive a Fokker-Planck equation for $P$ 
using standard procedures (see eg. [11]):
$$\partial _tP=(1/2)[\partial _{k_a}\partial _{k_b}D^{kk}_{ab}+2\partial _{k_a}
\partial _{B_b}D^{kB}_{ab}+\partial _{B_a}\partial _{B_b}D^{BB}_{ab}]P,\eqno(4)
$$
where the mean velocities vanish and the diffusivities are
$$D^{kk}_{ab}=\gamma [(d+1)k^2\delta _{ab}-2k_ak_b],\eqno(5)$$
$$D^{BB}_{ab}=\gamma [(d+1)B^2\delta _{ab}-2B_aB_b],\eqno(6)$$
$$D^{kB}_{ab}=-\gamma [(d+1)k_bB_a-k_aB_b-k_cB_c\delta _{ab}].\eqno(7)$$
For our purposes, it suffices to now the magnetic energy spectrum $W(k)$ 
defined as a $B^2$-moment of the probability distribution $P$. From (4-7), 
taking into account solenoidality $k_cB_c=0$, and assuming isotropy, one gets
the evolution equation for the magnetic energy spectrum 
$$\dot{W}=(d-1)k^2W''+(d^2-5)kW'+2(d^2-d-2)W,\eqno(8)$$
where prime stays for the modulus $k$ derivative, and time was normalized to 
exclude a factor proportional to $\gamma$. For $d=3$ this equation coincides 
with those obtained in [7,9]. 

Boundary conditions for (8) is the immediate question. First of all, molecular 
magnetic diffusivity will eat up all the modes with very large $k$. Thus, the 
boundary condition at $k=\infty $ is simply $W=0$. It is assumed that the limit $k_{max}\rightarrow \infty$ is taken after all other limits. The order of performing limits can be important. For example, magnetic energy $E=\int dk W$ in two dimensions without resistivity grows exponentially,
$$\lim_{t\rightarrow \infty }\lim_{k_{max}\rightarrow \infty }E=\infty .$$
At the same time, magnetic energy finally decays in two dimensions if the resistivity is positive, thus
$$\lim_{k_{max}\rightarrow \infty }\lim_{t\rightarrow \infty }E=0.$$
This follows from the Zeldovich antidynamo theorem, and our calculations will reproduce this result.

 Secondly, the very model 
(1),(2) is not applicable at small $k$, where the turbulence is not reducible 
to the corresponding rate of stretching. In fact, at large scales all we have 
is a usual turbulent magnetic diffusivity. The magnetic field is thrown out 
of the large-scale region into turbulent scales. The corresponding boundary 
condition should be $W'=0$ at some $k=k_{min}$. From what follows and because
$k_{min}$ is small, this boundary condition is equivalent to $W=0$ at $k=0$.

 The eigenmode of (8) is simply $k^y$, and the growth rate is 
$$\gamma (y)=(d-1)y(y-1)+(d^2-5)y+2(d^2-d-2).\eqno(9)$$
To satisfy the boundary conditions, one has to have a pair of solutions with 
the exponent $y=Rey\pm i0$ giving the same growth rate $\gamma$ (then a linear 
combination of these modes will vanish at $logk=\pm \infty$). This means that 
$y$ is determined by the requirement 
$$\gamma '(y)=0.\eqno(10)$$
 Calculating $y$ and then calculating the corresponding growth rate, we get 
the SSF dynamo ($\gamma >0$) for dimensionalities $d$ satisfying
$$d(d-1)(9-d)>16,\eqno(11)$$
or $2.103<d<8.765$ as advertised.

Certainly we are mostly interested in $d=3$ case, which happens to be a dynamo 
case according to (11). However, working in $d$ dimensions  is useful for what 
is coming. Suppose, that the $\delta$-correlation supposition is dropped. 
Then the coefficients in the diffusion equation (8), and in fact the very form 
of the equation will change. Perhaps we will still have SSF dynamo in some 
dimensionalities interval, but the lower critical dimension might turn out to 
be greater than 3, and there will be no SSF dynamo in three dimensions. As 
Vainshtein [12] puts it, the SSF dynamo is a quantitative rather than a 
qualitative problem. It was our original idea to show that the critical 
dimensions do depend on the correlation properties of turbulence, and the 
$\delta$-correlated result of [7,9] is not reliable. To our surprise, we 
found that the critical dimensions are structurally stable and probably even 
universal as far as the turbulence is incompressible and isotropic. We give a 
short account of our investigations in what follows.

The idea is to solve (1),(2) for a long rather than a short time interval. 
The solution is
$$B=UB_0,\eqno(12)$$
$$k=(U^t)^{-1}k_0.\eqno(13)$$
Here $U$ is the evolution matrix given by $\dot{U}=VU$, $U(t=0)=I$; in terms 
of Euler ${\bf r}$ and Lagrange ${\bf a}$ coordinates $U=\partial {\bf r}/
\partial {\bf a}$. If time of integration is long enough, the matrix $U$ can 
be written as $R_1DR_2$, where $R$ are random rotation matrices, and $D$ is 
diagonal. Physically speaking, after several correlation times, the orientation
of the fluid element is randomized. Now suppose 
that at time t=0 the magnetic energy spectrum was isotropic, $W_0=W_0(k)$, 
where $k$ now means the magnitude. Using randomness of the rotation matrices 
$R$, one can calculate the new spectrum $W(k)$ after the transformation (12),
(13) as
$$W(k)=<|D\hat{k}||D^{-1}\hat{b}|^{-(d+1)}W_0(|D\hat{k}|k)>, \eqno(14)$$
where $<...>$ means averaging over the angles of the perpendicular 
$d$-dimensional unit vectors $\hat{k}$ and $\hat{b}$. The integral 
transformation (14) generalizes the Fokker-Planck equation (8). Equation (14) 
is the main result, it can be derived as follows.

If $P_0$ was the probability distribution of pairs $({\bf k},{\bf B})$ before 
the transformation (12),(13), then the new pdf $P$ after the transformation is 
$$P({\bf k},{\bf B})=P_0(U^t{\bf k},U^{-1}{\bf B}).\eqno(15)$$
Because of isotropy, the magnetic energy spectrum $W(k)$ can be calculated as
$$W(k)=\int B^{d+1}dB<P(k\hat{k},B\hat{b})>,\eqno(16)$$
where $<...>$ is the angle average. Now plug (15) into (16)
$$W(k)=\int B^{d+1}dB<P_0(kU^t\hat{k},BU^{-1}\hat{b})>.\eqno(17)$$
As a consequence of isotropy and solenoidality, the probability distribution 
has a form
$$P_0({\bf k},{\bf B})=Q(k,B)\delta (\hat{k}\cdot \hat{b}),\eqno(18)$$
and (17) can be written as
$$W(k)=\int B^{d+1}dB<Q(|U^t\hat{k}|k,|U^{-1}\hat{b}|B)\delta (\hat{k}\cdot 
\hat{b}/(|U^t\hat{k}||U^{-1}\hat{b}|))>.\eqno(19)$$
Now change the integration variable in (19), $B\rightarrow B/|U^{-1}\hat{b}|$ 
and get
$$W(k)=\int B^{d+1}dB<|U^t\hat{k}||U^{-1}\hat{b}|^{-(d+1)}Q(|U^t\hat{k}|k,B)
\delta (\hat{k}\cdot \hat{b})>.\eqno(20)$$
Recalling (18) and the definition of $W_0$, we write (20) in a form (14). 
Only the diagonal part of $U$ matters because of the angle averaging.

 A nice property of (14) is that the eigenmode is still $k^y$. The 
amplification factor after a single transformation (14) is given by
$$\Gamma (y)=<|D\hat{k}|^{(y+1)}|D^{-1}\hat{b}|^{-(d+1)}>. \eqno(21)$$
Write the diagonal part as $D=exp~diag (\lambda _1/2,...,\lambda _d/2)$, the incompressibility implies $\lambda _1+...+\lambda _d=0$.
Then, 
$$|D\hat{k}|^2=\sum _je^{\lambda _j}k_j^2, \eqno(22)$$
$$|D^{-1}\hat{b}|^2=\sum _je^{-\lambda _j}b_j^2. \eqno(23)$$
Like before, to satisfy the boundary conditions one chooses the exponent $y$ 
in 
such a way as to minimize $\Gamma$:
$$\Gamma '(y)=0. \eqno(24)$$
Thus, the necessary and sufficient condition for the SSF dynamo is that 
$\Gamma$ be greater than 1 for arbitrary $y$ and $\lambda _j$. Since $\Gamma$ 
is equal to 1 at $\lambda _1=...=\lambda _d=0$, this point should be a minimum 
for the dynamo case. Calculating (21) in the vicinity of zero, one gets
$$\Gamma (y)=1+[4d(d-1)(d+2)]^{-1}(\sum _j\lambda _j^2)[(d-1)y(y-1)+(d^2-5)y+2(
d^2-d-2)].\eqno(25)$$
Comparing to (9), we see that $\Gamma$ is greater than 1 in the vicinity of of 
the point $\lambda =0$ if (11) is satisfied. Thus the structural stability of 
the result (11) is proved. We suspect that (11) ensures that $\Gamma$ is 
greater than 1 everywhere, not only in the vicinity of the zero point. We do 
not have a proof. However, numerical calculations of the integral (21) in 
three dimensions confirm that $\lambda =0$ is a global minimum and $\Gamma $ 
is greater than 1 everywhere else.

To conclude, we have shown that SSF dynamo operates in three dimensional 
isotropic turbulence. The result seems to be model independent and is 
definitely structurally stable. The SSF dynamo is of great importance to the
understanding of the growth of large scale fields. Galaxies and upper layers of
stars are characterized by $R_m\gg Re$ which is the necessary condition for a
SSF dynamo to be present.  The large growth rate of the SSF dynamo - it is
$Re^{1/2}$ bigger than the mean field growth rate - ensures that self
consistency effects first enter at the small scales, when the large scale
fields are negligible [9]. Clearly large scale fields are observed in galaxies
and stars, and the dynamo process must involve considerable evolution where the
small scale fields are dynamically important but the large scales are not.The 
SSF dynamo saturation mechanisms and the effects  of SSF dynamo on the 
conventional mean field dynamo should be subjects of further studies [9,13].

\vskip 0.3in
\noindent{\bf Acknowledgements}
\vskip 0.2in

It is a pleasure to thank Russell Kulsrud  for stimulating 
discussions and the referee for his/er very interesting questions.  This research was partially supported by the NASA Grant NAGW 2418,
 the D.O.E. Grant DE-FG03-88ER-53275, the NSF Grant PHY94-07194, and the NSF Grant PHY95-13835.

\vfill
\eject

\noindent{\bf References}
\vskip 0.2in
\parskip=0pt

\item {1.} H.K.Moffat, {\it Magnetic field generation in electrically 
conducting fluids}, (Cambridge University Press, 1978).
\item {2.} F.Cattaneo, S.I.Vainshtein, Astrophys.J. {\bf 376}, L21 (1991).
\item {3.} A.V.Gruzinov, P.H.Diamond, Phys.Rev.Lett. {\bf 72}, 1651 (1994)
\item {4.} A.V.Gruzinov, P.H.Diamond, Phys. Plasmas {\bf 2}, 1941 (1995)
\item {5.} G.K.Batchelor, Proc. Roy. Soc. {\bf A201},405 (1950)
\item {6.} R.Kraichnan, S.Nagarajan, Phys.Fluids {\bf 10}, 859 (1967)
\item {7.} A.P.Kazantsev, Sov.Phys.JETP {\bf 26}, 1031 (1968)
\item {8.} H.K.Moffatt, Rep.Prog.Phys. {\bf 46}, 621 (1983)
\item {9.} R.M.Kulsrud, S.W.Anderson, Astrophys.J. {\bf 396}, 606 (1992)
\item {10.} Ya.B.Zeldovich, A.A.Ruzmaikin, S.A.Molchanov, D.D.Sokoloff, 
J.Fluid. Mech. {\bf 144}, 1 (1984)
\item {11.} J.Zinn-Justin, {\it Quantum field theory and critical phenomena}, 
Chapter 4, (Clarendon Press, Oxford , 1993)
\item {12.} S.I.Vainshtein, Sov.Phys.JETP {\bf 52}, 1099 (1980)
\item {13.} E.G.Blackman, G.B.Field, "A new derivation and discussion of the
mean field dynamo equation", submitted to Phys.Rev.Lett. (1995)

\vfill
\eject

\end